\documentclass[%
 reprint,
 groupedaddress,
 amsmath,amssymb,
 aps,
 prl,
]{revtex4-2}

\usepackage{tikz}
\usepackage{graphicx}
\usepackage{dcolumn}
\usepackage{bm}
\usepackage{hyperref}
\usepackage{mathrsfs}

\usepackage{xcolor}
\usepackage{enumitem}
\usepackage{eso-pic}
\usepackage{braket}

\newcommand{\be}{\begin{equation}}
\newcommand{\ee}{\end{equation}}
\newcommand{\ba}{\begin{aligned}}
\newcommand{\ea}{\end{aligned}}
\newcommand{\bea}{\begin{eqnarray}}
\newcommand{\eea}{\end{eqnarray}}

\def\Tr{\mathop{\mathrm{Tr}}\nolimits}

\newcommand{\reportnum}[2]{
  \AddToShipoutPictureBG*{%
    \AtPageUpperLeft{%
      \hspace{0.75\paperwidth}%
      \raisebox{#1\baselineskip}{%
        \makebox[0pt][l]{\textnormal{#2}}
  }}}%
}

\def\mb{\mathbb}

\def\mc{\mathcal}
\def\bp{\begin{pmatrix}}
\def\ep{\end{pmatrix}}

\def\Tr{\mathop{\mathrm{Tr}}\nolimits}

\begin{document}

\reportnum{-3}{USTC-ICTS/PCFT-25-12}

\title{Symmetry Topological Field Theory for Flavor Symmetry}

\author{Qiang Jia$^1$}
\author{Ran Luo$^2$}
\author{Jiahua Tian$^3$}
\author{Yi-Nan Wang$^{2,4,5}$}
\author{Yi Zhang$^6$}

\affiliation{%
 $^1$Department of Physics, Korea Advanced Institute of Science \& Technology, Daejeon 34141, Korea
}
\affiliation{%
 $^2$School of Physics, Peking University 
}%
\affiliation{%
 $^3$School of Physics and Electronic Science, East China Normal University, Shanghai, China, 200241
}
\affiliation{$^4$Center for High Energy Physics, Peking University}

\affiliation{$^5$Peng Huanwu Center for Fundamental Theory, Hefei, Anhui 230026, China}

\affiliation{$^6$Kavli IPMU (WPI), UTIAS, The University of Tokyo, Kashiwa, Chiba 277-8583, Japan}

\date{\today}

\begin{abstract}
	In this Letter, we demonstrate that the Symmetry Topological Field Theory (SymTFT) associated to a Quantum Field Theory (QFT) with continuous non-abelian $G$-flavor symmetry is a $BF$-theory with gauge group $G$. We show that gauging $G$-symmetry with a flat connection yields a theory with global symmetry characterized by exchanging the conjugate variables in the quantization of $BF$-theory. We construct the extended operators that generate the $G$-flavor symmetry and the $(d-2)$-form $\text{Rep}(G)$-symmetry of the gauged QFT. We demonstrate that $BF$-theory arises as the theory characterizing $G$-flavor symmetry of a QFT in the AdS/CFT setup. 't Hooft anomalies of the $G$-flavor symmetry are realized as extra terms in the action.

\end{abstract}


\maketitle


\paragraph{Introduction.}

\begin{figure}[h]
	\includegraphics[width=0.5\textwidth]{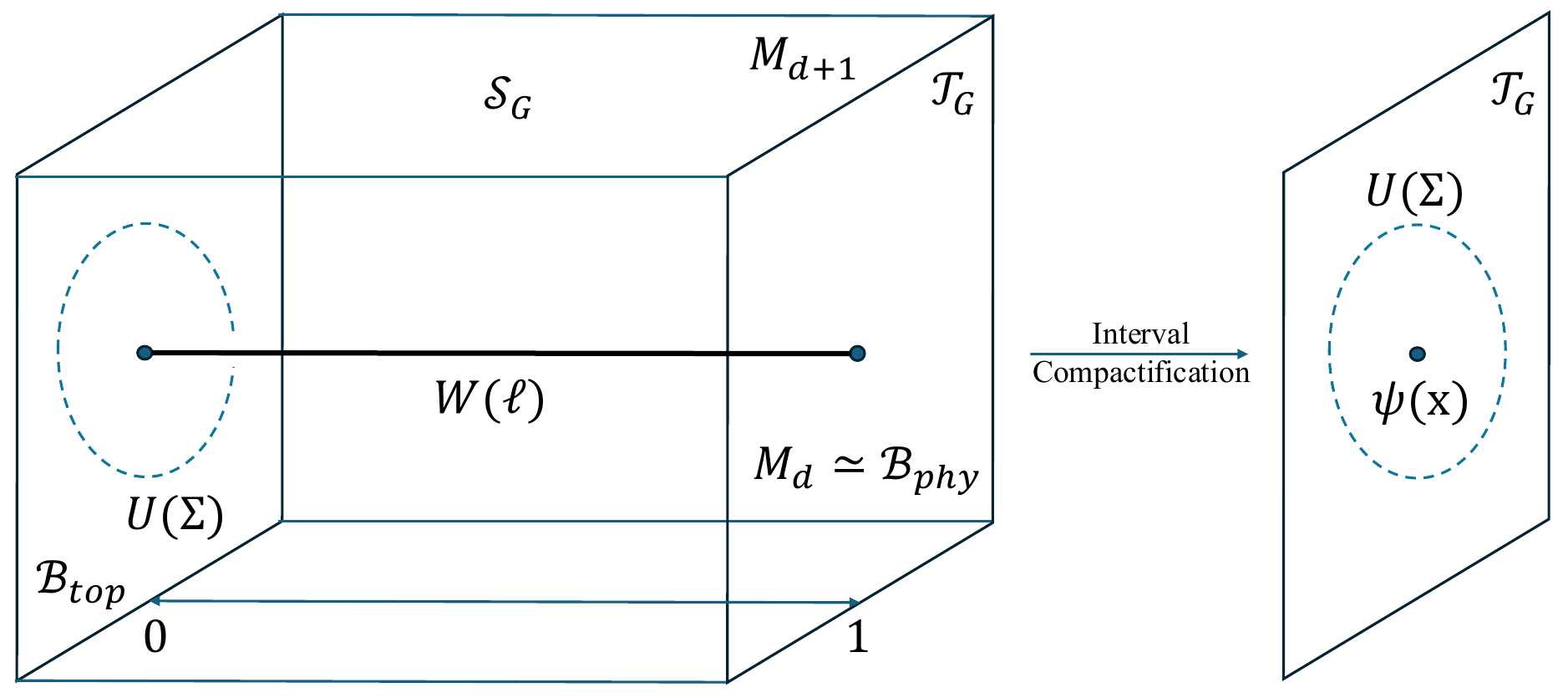}
	\caption{The relation between the SymTFT $\mathcal{S}_G$ and the corresponding physical theory $\mathcal{T}_G$. Various boundary conditions are imposed to linked operators in $\mathcal{S}_G$, whose interaction yields the $G$-symmetry action of $\mathcal{T}_G$ after interval compactification.}
	\label{fig:Sandwich}
\end{figure}

Since the notion of \emph{generalized global symmetry} of quantum field theory (QFT) was initially developed in~\cite{Alford:1990fc, Alford:1991vr, Alford:1992yx, NUSSINOV2009977, Kapustin:2013uxa, Kapustin:2014gua} then coined~\cite{Gaiotto:2014kfa}, tremendous efforts have been dedicated to studying this subject~\cite{Kaidi:2021xfk, Kaidi:2022cpf, Morrison:2020ool, Bhardwaj:2020phs, Bhardwaj:2022yxj, Bhardwaj:2023wzd, Bhardwaj:2023ayw,Bhardwaj:2024qiv,Bhardwaj:2025piv} with a central focus on understanding various gaugings of the generalized symmetries and their anomalies~\cite{Cordova:2019bsd, Cordova:2019jqi, Bhardwaj:2022dyt, Burnell:2021reh}, for which \emph{symmetry topological field theory} (SymTFT) has been a useful unifying framework~\cite{Witten:1998wy, Kong:2020cie, Gaiotto:2020iye, Apruzzi:2021nmk, Kaidi:2022cpf, vanBeest:2022fss, Freed:2022qnc, Apruzzi:2022dlm, Kaidi:2023maf, Baume:2023kkf, Cvetic:2024dzu}. The SymTFT $\mathcal{S}_G$ associated to a QFT $\mathcal{T}_G$ on $M_d$ of $G$-symmetry lives on $M_{d+1} \cong M_d\times [0,1]$ with topological boundary $\mathcal{B}_{\text{top}}$ at $x^{d+1}=0$ and physical boundary $\mathcal{B}_{\text{phy}}$ at $x^{d+1}=1$, as illustrated in Figure~\ref{fig:Sandwich}.
Various gaugings of $\mathcal{T}_G$ correspond to consistently imposing Dirichlet boundary conditions (DBC) or Neumann boundary conditions (NBC) to extended operators in $\mathcal{S}_G$. After an interval compactification, the interaction between linked extended operators yields the $G$-symmetry action of $\mathcal{T}_G$. It is usually tacitly assumed that these extended operators are topological, meaning they remain unchanged under any deformation unless it causes them to cross each other.

Although it was proposed in~\cite{Brennan:2024fgj, Antinucci:2024zjp, Bonetti:2024cjk} that the SymTFT of a QFT $\mathcal{T}_G$ with $G$-flavor symmetry is a $BF$-theory with gauge group $G$, a proof has yet to be provided. We will present concrete derivations to establish this proposal. More importantly, there still lacks a construction of the extended operators which yields the $G$-flavor symmetry action:
\begin{equation}\label{eq:nonab_law}
	U_g(\Sigma) \psi_i(x) = R(g)_{ij} \psi_j(x),\ g \in G
\end{equation}
for $U_g(\Sigma)$ supported on $\Sigma$ and local operator $\psi(x)$ in representation $R$ of $G$ when $\Sigma$ links $x$ in $\mathcal{T}_{G}$. It is rather easy to see the difficulty from Figure~\ref{fig:Sandwich}: to obtain $U(\Sigma)\psi(x)$ after interval compactification, one must consider the interaction between $U(\Sigma)$ and a line operator $W(\ell)$ ending on $\psi(x)$ in $\mathcal{S}_G$; however, no line operators can transform in any non-abelian fashion as in~(\ref{eq:nonab_law}) under the action of a linked topological operator $U(\Sigma)$~\cite{Gaiotto:2014kfa}. We will resolve this conundrum by allowing non-topological extended operators in $\mathcal{S}_G$.

In this Letter, following the construction in~\cite{Witten:1998wy, Witten:2009at}, we define the SymTFT $\mathcal{S}_G$ of a theory $\mathcal{T}_G$ on $M_d$ to be a topological field theory $\mathcal{S}_G$ on $M_{d+1}\cong M_d\times [0,1]$, whose canonical quantization over ``time-slice'' $M_d$ yields a state space that coincides with the space of partition functions of $\mathcal{T}_G$ minimally coupled to background flat $G$-gauge field. We then derive that $\mathcal{S}_G$ is the $BF$-theory with gauge group $G$. With our definition, the equivalence of gauging and Fourier transform, as observed in~\cite{Witten:1998wy, Gaiotto:2014kfa}, becomes transparent and is used to obtain the spaces that parametrize the operators in $\mathcal{S}_G$ that generate either the $G$-flavor symmetry of $\mathcal{T}_G$ or the dual symmetry of the gauged theory on $M_d$. Quite surprisingly, unlike most known cases, we find that those operators do not have to be topological, nevertheless, they yield~(\ref{eq:nonab_law}) on $M_d$ when they link in $M_{d+1}$ with proper boundary conditions. This indicates the necessity of looking at a broader spectrum of extended operators in general SymTFT. We present an example of $G=SU(2)$ and classify all possible topological boundary conditions and the resulting global symmetries. We show that $BF$-theory can also be derived as the SymTFT of a QFT living on the boundary in a holographic setup. Lastly, we discuss the incorporation of 't Hooft anomalies in $\mathcal{S}_G$ that obstruct gaugings.

\paragraph{Partition functions from quantization of $BF$-theory.}

In the absence of 't Hooft anomaly, gauging continuous $G$-symmetry of $\mathcal{T}_G$ is done schematically by coupling $\mathcal{T}_G$ to flat gauge field $A$ as~\cite{Witten:1998wy, Kaidi:2021xfk, Kaidi:2022cpf, Bhardwaj:2023ayw, Bhardwaj:2023kri, Schafer-Nameki:2023jdn}:
\begin{equation}\label{eq:continuous_gauging}
    Z_{\mathcal{T}_G/G}[B'] = \frac{1}{\text{Vol}(\mathcal{G})} \int \mathcal{D}A\ Z_{\mathcal{T}_G}[A] \exp\left( i \langle A, B' \rangle_{M_d} \right)\,.
\end{equation}
Here $\mathcal{G}$ is the space of all gauge transformations of principal $G$-bundle $P(G,M_d)$ on $M_d$, $\langle A, B' \rangle_{M_d}$ is a non-degenerate gauge-invariant pairing on $M_d$ of flat connection $A$ of $P(G,M_d)$ and some dual field $B'$, and $Z_{\mathcal{T}_G}[A]$ is the partition function of $\mathcal{T}_G$ minimally coupled to $A$ via $A\wedge J$ for conserved current $J$ and $Z_{\mathcal{T}_G/G}[B']$ is the partition function of the gauged theory $\mathcal{T}_G/G$ coupled to $B'$. A sum over inequivalent principal $G$-bundles in~(\ref{eq:continuous_gauging}) is implicitly assumed. Self-consistency of~(\ref{eq:continuous_gauging}) requires $Z_{\mathcal{T}_G}[A]$ be invariant under gauge transformations of $A$. In other words, $Z_{\mathcal{T}_G}[A]$ must be a function on $\widetilde{\mathcal{A}}:= \mathcal{A}_{\text{flat}}/\mathcal{G}$ where $\mathcal{A}_{\text{flat}}$ is the space of all flat connections of $P(G,M_d)$.

It has long been known that $\widetilde{\mathcal{A}}$ is the space of classical solutions of non-abelian $BF$-theory defined on $M_{d+1} \cong M_d\times \mathbb{R}$ with action~\cite{Horowitz:1989ng, Blau:1989bq}:
\begin{equation}\label{eq:BF_action}
    S_G = \int_{M_{d+1}} \text{Tr}B\wedge F \equiv \int_{M_{d+1}} \text{Tr}B\wedge D_AA
\end{equation}
for $D_A = d -i A\wedge$ with $A$ a connection of $P(G,M_{d+1})$ and $\text{ad}(P)$-valued $(d-1)$-form $B$. The action is invariant under both $G$-gauge transformations $A \rightarrow g A g^{-1} + ig d g^{-1}$ and $B\rightarrow g B g^{-1}$ with $g\in\mathcal{G}$, and a shift $B \rightarrow B + D_AK$ with $\text{ad}(P)$-valued $(d-2)$-form $K$.
The canonical quantization of~(\ref{eq:BF_action}) over time-slice $M_d$ yields wave functions living in the state space $\text{Fun}(\widetilde{\mathcal{A}})$, the space of functions on $\widetilde{\mathcal{A}}$~\cite{Baez:1999sr}.

The above discussions suggest viewing $Z_{\mathcal{T}_\mathcal{G}}[A]$ as a wave function of the quantum mechanical system defined by~(\ref{eq:BF_action}), i.e. $Z_{\mathcal{T}_\mathcal{G}}[A] \in \text{Fun}(\widetilde{\mathcal{A}})$. Therefore the SymTFT $\mathcal{S}_G$ of $\mathcal{T}_G$ must be the $BF$-theory with action~(\ref{eq:BF_action}) by our definition, since its state space coincides with the space of $Z_{\mathcal{T}_\mathcal{G}}[A]$. We will substantiate this one-line proof in the subsequent sections. We emphasize that since $B$ in~(\ref{eq:BF_action}) is not the dual field $B'$ in~(\ref{eq:continuous_gauging}), $BF$-theory is not a conventional SymTFT as in e.g.~\cite{Apruzzi:2021nmk, Kaidi:2021xfk, Kaidi:2023maf, Bhardwaj:2023ayw, Schafer-Nameki:2023jdn}.

\paragraph{Gauging as Fourier transform.}

Viewing $Z_{\mathcal{T}_\mathcal{G}}[A]$ as a wave function of the quantum mechanical system defined by~(\ref{eq:BF_action}) has an extra advantage that it enables us to view~(\ref{eq:continuous_gauging}) as changing the polarization of quantization of $\mathcal{S}_G$. To this end, recall that for path-connected $M_d$ we have $\widetilde{\mathcal{A}} \cong \text{Hom}(\pi_1(M_d), G)/G$ where $G$ acts on $\phi\in \text{Hom}(\pi_1(M_d), G)$ by $g\phi(\gamma)g^{-1}$ for $\gamma\in \pi_1(M_d)$~\cite{Kobayashi&Nomizu}. For simplicity, we assume $\pi_1(M_d) = \mathbb{Z}$ in which case $\widetilde{\mathcal{A}} = Cl(G)$, the set of conjugacy classes of $G$. By Tannaka duality~\cite{Tannaka, Tatsuuma} there exists an isometry~\cite{IntroTannaka}:
\begin{equation}\label{eq:Fourier_trans}
    L^2(Cl(G)) \stackrel{\mathsf{T}}{\cong} L^2(G^\vee)
\end{equation}
via Fourier transform $\mathsf{T}$ where $G^\vee$ is the isomorphism classes of irreducible representations of $G$~\footnote{See SM1 for a short review of useful facts about Tannaka duality.}. Therefore~(\ref{eq:continuous_gauging}) as a Fourier transform must be equivalent to $\mathsf{T}$ which exchanges the conjugate variables in the canonical quantization of $\mathcal{S}_G$~\cite{bates1997lectures}. Physically,~(\ref{eq:Fourier_trans}) implies that inequivalent $B'$ must live in $G^\vee$ when $\pi_1(M_d) = \mathbb{Z}$.

A more physical implication of~(\ref{eq:Fourier_trans}) is that when $\pi_1(M_d) = \mathbb{Z}$, the parameter space of $G$-symmetry generators of $\mathcal{T}_G$ must be a certain fibration over $Cl(G)$ while that of the symmetry generators of $\mathcal{T}_G/G$ must be a certain fibration over $G^\vee$. Later we will construct these generators and verify they indeed satisfy such implication.

\paragraph{Extended operators in $BF$-theory.}

Having shown that the quantization of $BF$-theory reproduces the space of partition vectors of $\mathcal{T}_G$ and encodes gauging as switching representations via Fourier transform, we go on to the construction of the extended operators in the $BF$-theory that later will be found to yield the $G$-action~(\ref{eq:nonab_law}) of $\mathcal{T}_G$ after interval compactification.

On one hand, we consider an untraced Wilson line in representation $\mathbf{R}$ stretching between fixed points of $\mathcal{B}_{\text{top}}$ and $\mathcal{B}_{\text{phy}}$:
\be\label{eq:W_op_main}
W_{\mathbf{R}}(\ell) = \mathcal{P} \exp\left( i\int_\ell A_\mathbf{R} \right)\,.
\ee
where $A_\mathbf{R}$ is in representation $\mathbf{R}$ of $G$. On the other hand, for any covariantly constant section $\alpha$ of $\text{ad}(P)$ with a flat connection, i.e. $D_A\alpha = 0$ for certain flat $A$, we define the following operator:
\be
\label{eq:Uop_main}
U_\alpha(\Sigma) = \exp\left(i \int_\Sigma(\alpha,B)\right)
\ee
where $\Sigma$ is a codimension-2 submanifold of $M_{d+1}$ and $(\cdot,\cdot)$ is an ad-invariant inner product of $\mathfrak{g}$.

However, $U_\alpha(\Sigma)$ is not topological. To see this, one can compute $U_\alpha(\Sigma) W_\mathbf{R}(\ell)$ for which we have~\footnote{See SM2 for the details.}:
\begin{equation}\label{eq:UW_nontopo}
    \begin{split}
        U_\alpha(\Sigma) W_\mathbf{R}(\ell) &= \mathcal{P} \left [e^{i\int_p^1 A_\mathbf{R}} e^{-i\alpha_\mathbf{R}(p)} e^{i\int_0^p A_\mathbf{R}} \right]
    \end{split}
\end{equation}
when $\Sigma$ is the boundary of an infinitesimal $d$-ball that intersects $\ell$ parametrized by $[0,1]$ transversally at $p$. Since $W_{\mathbf{R}}(\ell)$ is topological while~(\ref{eq:UW_nontopo}) is not due to its explicit $p$-dependence, $U_\alpha(\Sigma)$ cannot be topological. Actually, no non-commutative symmetry can be generated by codimension-2 topological operators~\cite{Gaiotto:2014kfa}, and~(\ref{eq:UW_nontopo}) is nothing but a use case of this fact.

There are a few ways to obtain $p$-independent actions from~(\ref{eq:UW_nontopo}). The simplest is to let $e^{-i\alpha_\mathbf{R}(p)}$ be an element of the center $Z(G)$ of $G$ to make $U_\alpha(\Sigma)$ a generator of the 1-form $Z(G)$-symmetry of $BF$-theory. Another is to define an operator $\widetilde{U}_{\widetilde{\alpha}}(\Sigma)$ for a fixed flat connection $A$:
\begin{equation}\label{eq:Utilde}
    \widetilde{U}_{\widetilde{\alpha}}(\Sigma) = \int dg \exp\left( i\int_\Sigma(\widetilde{\alpha}, B) \right)
\end{equation}
where $dg$ is the Haar measure of $G$ and $\widetilde{\alpha}$ is the parallel transport of $g\alpha(p)g^{-1}$ for a covariantly constant $\alpha$ at $p$ by the holonomy of $A$ on each local patch. The above definition is independent of the choice of $p$. Unlike $\alpha$, we allow $\widetilde{\alpha}$ to be defined up to conjugation. In other words, $e^{-i\widetilde{\alpha}}\in Cl(G)$. The action of $\widetilde{U}_{\widetilde{\alpha}}(\Sigma)$ on $W_\mathbf{R}(\ell)$ is~\footnote{See SM2 for the details.}:
\begin{equation}\label{eq:UtildeW_pre}
    \begin{split}
        \langle \widetilde{U}_{\tilde{\alpha}}(\Sigma) W_\mathbf{R}(\ell) \rangle &= \mathcal{P} \left[ e^{i\int_p^1 A_\mathbf{R}}\ \int dg g e^{-i\alpha_\mathbf{R}(p)} g^{-1}\ e^{i\int_0^p A_\mathbf{R}} \right] \\
        &= \frac{\chi_\mathbf{R}(e^{-i\widetilde{\alpha}})}{\dim \mathbf{R}} \langle W_\mathbf{R}(\ell) \rangle
    \end{split}
\end{equation}
for irreducible representation $\mathbf{R}$ and Schur's lemma is applied to obtain the second line~\cite{Cordova:2022rer}. Here $\frac{\chi_\mathbf{R}(e^{-i\widetilde{\alpha}})}{\dim \mathbf{R}}$ is the \emph{normalized character}, which forms a basis of $L^2(Cl(G))$. Consequently, $\langle \widetilde{U}_{\widetilde{\alpha}} (\Sigma) W_\mathbf{R}(\ell) \rangle$ is topological.

\paragraph{Flavor symmetry and dual symmetry from $BF$-theory.}

\begin{figure}[h]
    \centering
    \includegraphics[width=0.9\linewidth]{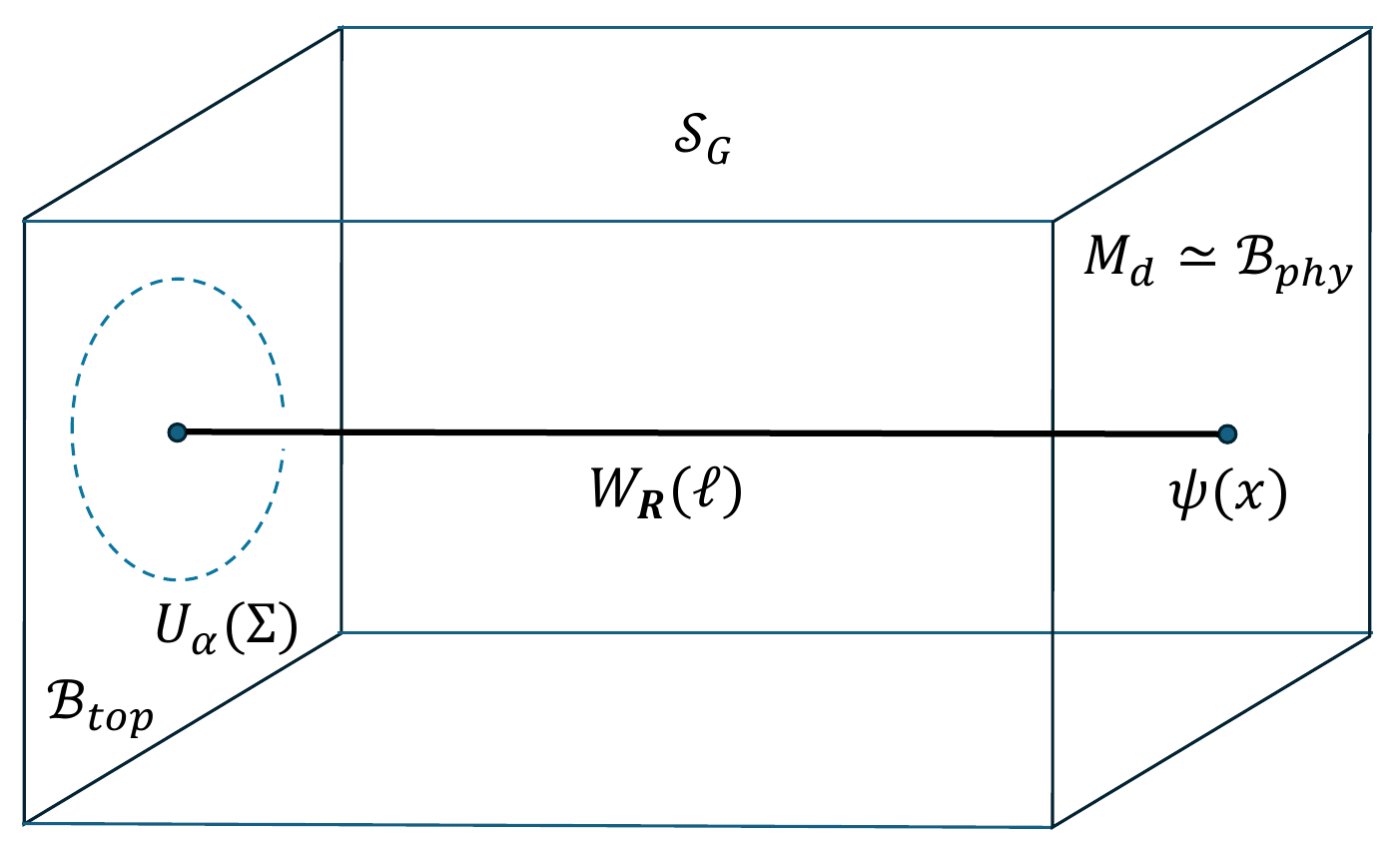}
    \caption{$U_\alpha(\Sigma)$ acts on the composite operator $\psi(x)W_{\mathbf{R}}(\ell)$.}
    \label{fig:U_on_boundary}
\end{figure}

Until now it may seem somewhat unclear why we have considered a non-topological action~(\ref{eq:UW_nontopo}) in a topological field theory, however we will show that it leads to the desired non-abelian symmetry action~(\ref{eq:nonab_law}). To see this, upon fixing $p = 0$,~(\ref{eq:UW_nontopo}) becomes:
\begin{equation}\label{eq:UW_fixed_p}
    U_\alpha(\Sigma) W_\mathbf{R}(\ell) = \mc{P}\left[W_{\mathbf{R}}(\ell) e^{-i\alpha_\mathbf{R}(0)}\right]\,.
\end{equation}
We see that not only is the $p$-dependence of $U_{\alpha}(\Sigma) W_{\mathbf{R}}(\ell)$ turned off manually, but also we observe a striking similarity between~(\ref{eq:UW_fixed_p}) and~(\ref{eq:nonab_law}). Actually, physically to let $W_{\mathbf{R}}(\ell)$ stretch between $\mathcal{B}_{\text{top}}$ and $\mathcal{B}_{\text{phy}}$, one has to impose DBC on $A$ at $\mathcal{B}_{\text{top}}$ and put a local operator $\psi(x)$ at $\mathcal{B}_{\text{phy}}$ on which $W_{\mathbf{R}}(\ell)$ ends as illustrated in Figure~\ref{fig:U_on_boundary}.
Therefore, instead of~(\ref{eq:UW_fixed_p}), physically $U_{\alpha}(\Sigma)$ acts on the composite operator $\psi(x)W_{\mathbf{R}}(\ell)$. Using~(\ref{eq:UW_nontopo}), we have:
\begin{equation}\label{eq:U_WQ_main}
    \begin{split}
        U_{\alpha}(\Sigma) \psi(x)W_\mathbf{R}(\ell) &= \lim_{p\rightarrow 0}\mathcal{P} \left[ \psi(x) e^{i\int_p^1 A_\mathbf{R}} e^{-i\alpha_\mathbf{R}(p)} e^{i\int_0^p A_\mathbf{R}} \right] \\
        &= \mc{P}\left[\psi(x)W_\mathbf{R}(\ell) e^{-i\alpha_\mathbf{R}(0)}\right]\,.
    \end{split}
\end{equation}
A compactification of $\mathcal{S}_G$ on $[0,1]$ recovers $\mathcal{T}_G$ by effectively sending the length of $\ell$ to zero~\cite{Witten:1998wy, Freed:2012bs, Apruzzi:2021nmk}. In this limit,~(\ref{eq:U_WQ_main}) becomes:
\begin{equation}\label{eq:psi_in_R}
\begin{split}
    U_{\alpha}(\Sigma) \psi(x) &= \mc{P}\left[\psi(x) e^{-i\alpha_{\mathbf{R}}(0)}\right]\cr
    &=e^{-i\alpha_{\mathbf{R}}(0)}\psi(x)
    \end{split}
\end{equation}
for $\Sigma$ linking $x$ in $\mathcal{B}_{\text{phy}}\cong M_d$. Therefore we reproduce the non-abelian symmetry action on a charged local operator of $\mathcal{T}_G$.

Exchanging boundary conditions of $U_\alpha(\Sigma)$ and of $W_\mathbf{R}(\ell)$ gauges $G$-symmetry, as a consequence of which $W_\mathbf{R}(\ell)$ generates the symmetry of $\mathcal{T}_G/G$ after imposing DBC on $B$ and NBC on $A$. Note that imposing NBC on $A$ then letting $\ell$ be a closed loop necessarily makes $W_\mathbf{R}(\ell)$ a traced Wilson loop and for simplicity we do not notationally distinguish an untraced Wilson line from a traced Wilson loop whenever the context is clear.

We see that the parameter spaces of~(\ref{eq:W_op_main}) and~(\ref{eq:Uop_main}) are indeed fibrations over the corresponding moduli spaces implied by~(\ref{eq:Fourier_trans}). Concretely, we again fix $\pi_1(M_d) = \mathbb{Z}$. On one hand, the parameter of a traced $W_\mathbf{R}(\ell)$ lives in a fibration over $G^\vee$ whose fiber at generic $\mathbf{R} \in G^\vee$ is $\mathbb{R}$ which parametrizes rescaling of $A$. On the other hand, the parameter $\alpha$ of $U_\alpha(\Sigma)$ lives in a fibration over $Cl(G)$ whose fiber at a generic $h\in Cl(G)$ is the Cartan subalgebra of $G$.

Traced $W_\mathbf{R}(\ell)$ as generators of symmetry of $\mathcal{T}_G/G$ fuse as:
\begin{equation}\label{eq:WR_fuse}
    W_{\mathbf{R}_1}(\ell)\times W_{\mathbf{R}_2}(\ell) = \sum_{i} W_{\mathbf{R}_{12i}}(\ell)
\end{equation}
for $\mathbf{R}_{12i}$ in the tensor product decomposition of $\mathbf{R}_1\otimes \mathbf{R}_2$. Therefore $\mathcal{T}_G/G$ has a $\text{Rep}(G)$ $(d-2)$-form symmetry, whose objects fuse the same way as~(\ref{eq:WR_fuse}), which is a non-invertible categorical symmetry~\cite{Frohlich:2004ef, Teo:2015xla, Barkeshli:2014cna, Bhardwaj:2017xup}.

\paragraph{Various gaugings of $\mathcal{T}_{G}$.}

We now study various gaugings of $\mathcal{T}_G$ using our prescription. To be concrete we fix $G=SU(2)$ and $\pi_1(M_d)=\mathbb{Z}$. For simplicity we use $\mathcal{G}^{(p)}$ to denote a $p$-form $\mathcal{G}$-symmetry where $\mathcal{G}$ is not necessarily a group. In this case $G^\vee\cong\frac{1}{2}\mathbb{N}$ and $\widetilde{\mathcal{A}}=Cl(G)\cong[0,\pi]$. The pairing between $j\in G^\vee$ and $\theta\in Cl(G)$ in Fourier transform~(\ref{eq:Fourier_trans}), as suggested by~(\ref{eq:UtildeW_pre}), is the normalized character~$\frac{\sin (2j+1)\theta}{(2j+1)\sin\theta}$ which furthermore is the Dirac pairing between topological operators with DBC. For a valid choice of boundary conditions at $\mathcal{B}_{\text{top}}$, the pairing between any two operators with DBC should be one~\cite{Kaidi:2022cpf}. Given this requirement, we find the following sets of boundary conditions at $\mathcal{B}_{\text{top}}$ and the global symmetry of the theory at $\mathcal{B}_{\text{phy}}$:

\begin{enumerate}
\item Impose DBC to all $W_j(\ell)$. All $U_\alpha(\Sigma)$ are with NBC and generate $SU(2)^{(0)}$ symmetry of the theory at $\mathcal{B}_{\text{phy}}$.

\item Impose NBC to all $W_j(\ell)$. They generate ${\rm Rep}(SU(2))^{(d-2)}$ symmetry of the theory at $\mathcal{B}_{\text{phy}}$.

\item Impose DBC to $W_j(\ell)$ with $j\in\mb{Z}$ and to $\widetilde{U}_0(\Sigma)$ and $\widetilde{U}_\pi(\Sigma)$. All other operators are with NBC and generate $SO(3)^{(0)}\times \mb{Z}_2^{(d-2)}$ symmetry of the theory at $\mathcal{B}_{\text{phy}}$. This is equivalent to gauging the $\mb{Z}_2$ center of the $SU(2)^{(0)}$ symmetry.

\end{enumerate}
These three sets of boundary conditions are related by gaugings as shown in Figure~\ref{fig:3-gaugings}.

\begin{widetext}
\begin{center}
\begin{figure}[htbp]
\includegraphics[height=2.5cm]{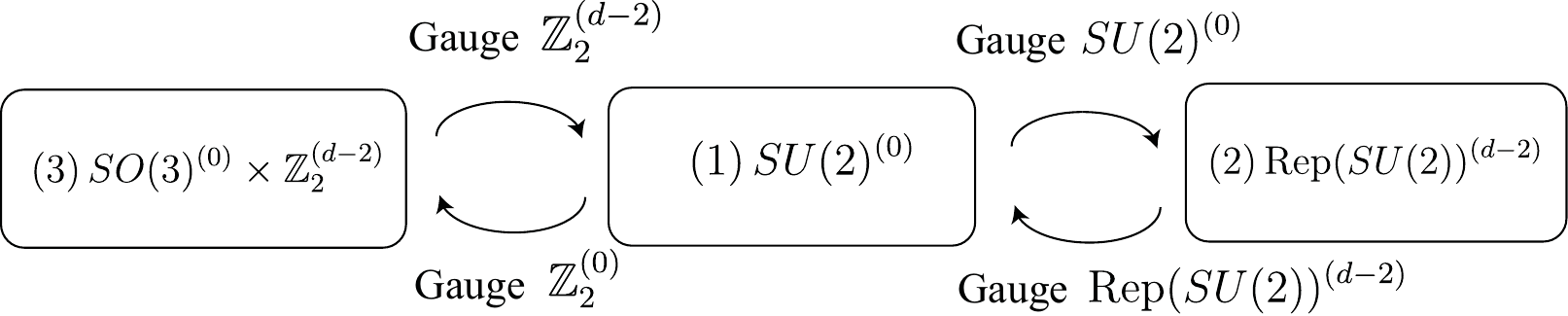}
\caption{The gaugings relating the three sets of boundary conditions for $G=SU(2)$.}\label{fig:3-gaugings}
\end{figure}
\end{center}
\end{widetext}

A boundary condition at $\mathcal{B}_{\text{top}}$ is equivalent to gauging $SO(3)$ in $\mathcal{T}_{SU(2)}$ does not exist. Such a gauging would have been possible if DBC were imposed to all $W_j(\ell)$ for half-integer $j$ and all $\tilde{U}_{\alpha}(\Sigma)$ for $\alpha \in (0,\pi)$. This clearly contradicts the requirement that the pairings between these operators must be one. To see this from a different point of view, note that~(\ref{eq:continuous_gauging}) implies that gauging $G'$-symmetry can be done by restricting the path integral to the space of flat $G'$-connections $\widetilde{\mathcal{A}}_{G'}$ which is consistent only if $\widetilde{\mathcal{A}}_{G'}$ is a closed subspace of $\widetilde{\mathcal{A}}$. Hence gauging $SO(3)$ of $\mathcal{T}_{SU(2)}$ is impossible since $\widetilde{\mathcal{A}}_{SO(3)} \cong Cl(SO(3))$ is a $\mathbb{Z}_2$ folding, rather than a closed subspace, of $[0,\pi]$. On the other hand, gauging $\mathbb{Z}_2$ in $\mathcal{T}_{SU(2)}$ is possible since $\widetilde{\mathcal{A}}_{\mathbb{Z}_2}\cong \{0,\pi\}$ is a closed subspace of $[0,\pi]$.

\paragraph{$BF$-theory from holography.}

Let us revisit the derivation leading to~(\ref{eq:psi_in_R}). Comparing with the holographic construction in~\cite{Harlow:2018jwu, Harlow:2018tng}, we see that $U_\alpha(\Sigma)$ is analogous to an \emph{asymptotic symmetry operator} acting on a Wilson line extending to the AdS boundary with \emph{long-range gauge symmetry} in the bulk. It is thus interesting to see if the analogy can be sharpened and if $BF$-theory arises in a holographic setup.

It is well-known in AdS$_{d+1}$/CFT$_d$~\cite{Maldacena:1997re, Witten:1998qj} that a covariantly conserved current $J^\mu$ at $M_d$ is the response of $\int_{M_d} d^dx A_\mu J^\mu$ to variations of gauge field $A_\mu$. Such minimal coupling can be obtained by calculating the holographically renormalized on-shell action~\cite{Bianchi:2001de, Bianchi:2001kw, Skenderis:2002wp} in a way that parallels the derivation in e.g.~\cite{Marolf:2006nd, DeWolfe:2020uzb}. For concreteness, we fix $d=4$ where one can find the variation of the renormalized action to be~\footnote{See SM3 for derivation.}:
\begin{equation}\label{eq:deltaStot_main}
    \delta {S}_{\text{ren}} = 2L^3 \int_{M_4} d^4x \Tr\delta\alpha_i\beta^i
\end{equation}
where $L$ is the radius of AdS$_5$ with boundary $M_4$. The current $\langle J^i \rangle \equiv 2L^3\beta^i$ on $M_4$ is thus covariantly conserved by the equation of motion $D_i\beta^i = 0$.

Physically, any bulk action defined in the neighbor of the boundary whose variation is identical to~(\ref{eq:deltaStot_main}) must lead to the same current. To this end, we find that the variation of on-shell $BF$-action on a space homeomorphic to $M_4\times (0,1]$ is:
\begin{equation}\label{eq:BF_variation_main}
    \delta S = \int_{M_4} d^4x \Tr \delta A_j (\widetilde{B}^{rj})
\end{equation}
where $\widetilde{B}$ is the Hodge dual of $B$. Note that~(\ref{eq:BF_variation_main}) takes the same form as~(\ref{eq:deltaStot_main}). By the equation of motion $D_i \widetilde{B}^{ri} = 0$, we immediately see that a covariantly conserved current $\langle J^i \rangle \equiv \widetilde{B}^{ri}$ arises from the variation~(\ref{eq:BF_variation_main}), which is physically equivalent to that obtained from Yang-Mills action on the same space. In this sense, we claim that $BF$-theory with gauge group $G$ can be derived to be the SymTFT of a theory on the boundary with $G$-flavor symmetry in the holographic setup.

Furthermore, identification of $\widetilde{B}^{ri}$ with the covariantly conserved current in a holographic setup leads to identifying~(\ref{eq:psi_in_R}) with the action of long-range gauge symmetry on the endpoint of a Wilson line described in~\cite{Harlow:2018jwu, Harlow:2018tng} (see~\cite{Heckman:2024oot, Cvetic:2025kdn} for relevant discussions). Moreover, since $BF$-theory $M_d\times (0,1]$ leads to the same consequences regarding flavor symmetry of the theory on $M_d$ as a theory with long-range gauge symmetry in the bulk, it must also share similar properties of the quantum gravity theory in AdS$_{d+1}$. To this end, note that $BF$-theory with gauge group $G$ has the following properties:
\begin{enumerate}
    \item There are no flavor symmetries;
    \item An untraced Wilson line can be in any representation of $G$;
    \item $G$ is compact.
\end{enumerate}
They align almost trivially with the following three main conjectures on symmetries in quantum gravity~\cite{Misner:1957mt, Polchinski:2003bq, Banks:2010zn}, when specialized to flavor or ordinary gauge symmetry:
\begin{enumerate}
    \item There are no global symmetries; 
    \item There are dynamical objects in all irreducible representations of internal gauge symmetry;
    \item The gauge group is compact.
\end{enumerate}

\paragraph{Twist terms and 't Hooft anomaly.}

So far, no obstruction to gauging either the $G^{(0)}$ or ${\rm Rep}(G)^{(d-2)}$ symmetry has been introduced. To address this, we can add a \emph{twist term} $I(A)$ or $I(B)$ to the action~(\ref{eq:BF_action}) to incorporate 't Hooft anomalies~\cite{Antinucci:2024bcm}. 

As an example, we consider the 5d SymTFT for 4d $SU(N_c)$ QCD with $N_f$ fundamental quarks, which has flavor symmetry $SU(N_f)_L\times SU(N_f)_R\times U(1)_V$. It is known that such flavor symmetry has a 't Hooft anomaly that obstructs its gauging. In our prescription, it is encoded in the action by adding a twist term $I(A)$ to~(\ref{eq:BF_action}), which typically is proportional to the Chern-Simons term with gauge group $SU(N_f)_L\times SU(N_f)_R\times U(1)_V$~\cite{Zumino:1983ew, Alvarez-Gaume:1984zlq, Harvey:2005it, Bilal:2008qx}.

Due to the presence of $I(A)$, it is impossible to impose NBC to $A$. For concreteness, let the twist term be $I(A_L)$ where $A_L$ is the background flat gauge field coupled to the conserved current of $SU(N_f)_L$. The variation of the on-shell action with respect to $A_L$ yields an additional boundary term $\int_{M_d\times\{0\}} f(A_L) \delta A_L$, where $f(A_L)$ is the response of $I(A_L)$ to the variation of $A_L$. Requiring the boundary term to vanish is incompatible with the NBC of $A_L$ since $f(A_L)$ can fluctuate on the boundary. In other words, it is impossible to gauge $SU(N_f)_L$-flavor symmetry in the presence of $I(A_L)$.

Moreover, the presence of twist term $I(A_L)$ obstructs the procedure of integrating out $B$ field to get the flatness condition $F=0$ in the bulk, as the bulk action itself is no longer gauge invariant under the gauge transformation of $A_L$. 

It will be interesting to further test our prescription against known cases with 't Hooft anomalies, e.g. 6d superconformal field theories~\cite{Heckman:2013pva, Ohmori:2014kda, Heckman:2015bfa, Heckman:2018jxk}. In particular, it is almost trivial to incorporate 't Hooft anomalies of continuous flavor symmetry or $R$-symmetry in our prescription of SymTFT, since it is a bottom-up approach that does not depend on any string compactification. To this end, we note that no known top-down stringy construction of the 7d SymTFT associated to a 6d SCFT~\cite{Witten:1998wy, Apruzzi:2022dlm, Lawrie:2023tdz, Tian:2024dgl} leads to twist terms that incorporate either continuous flavor symmetry or $R$-symmetry.

\vspace{0.3cm}
\begin{acknowledgments}

We would like to thank Chi-Ming Chang, Meng Cheng, Sergei Gukov, Yuting Hu, Ziming Ji, Satoshi Nawata, Benjamin Sung, Yuji Tachikawa, Zhenbin Yang and Yehao Zhou for discussions. We would like to thank the Peng Huanwu Center for Fundamental Theory for hosting the ``SymTFT workshop'' and the hospitality. JT would like to thank the hospitality of BIMSA where some motivations of this work were initially inspired. RL, YNW and YZ are supported by National Natural Science Foundation of China under Grant No. 12175004, No. 12422503 and by Young Elite Scientists Sponsorship Program by CAST (2024QNRC001). YNW is also supported by National Natural Science Foundation of China under Grant No. 12247103. QJ is supported by National Research Foundation of Korea (NRF) Grant No. RS-2024-00405629 and Jang Young-Sil Fellow Program at the Korea Advanced Institute of Science and Technology. JT is supported by National Natural Science Foundation of China under Grant No. 12405085 and by the Natural Science Foundation of Shanghai (Grant No. 24ZR1419300). YZ is supported by National Natural Science Foundation of China under Grant No. 12305077. Kavli IPMU is supported by World Premier International Research Center Initiative (WPI), MEXT, Japan.

\end{acknowledgments}


\bibliography{symtft}

\onecolumngrid

\appendix

\section{SM1: Fourier Transform and Tannaka Duality}
\paragraph{The abelian story.} Locally compact abelian groups, e.g. discrete abelian groups $\mathbb{Z}$, $\mathbb{Z}_n$, $\mathbb{Q}$ with discrete topology or $\mathbb{R}^n$ for any $n$ and $U(1)$ with standard topology, play significant roles in the study of generalized global symmetries. The Pontryagin dual $\widehat{G}$ of a locally compact abelian group $G$ is defined as the set of continuous group homomorphisms (characters $\chi$) from $G$ to $U(1)$
\begin{equation*}
    \widehat{G} = \text{Hom}(G,U(1))\,,
\end{equation*} 
which turns out to be locally compact, abelian and isomorphic to the set of isomorphic classes of complex irreducible representations of $G$. Gauging a theory with global symmetry $G$ leads to another theory with dual symmetry $\widehat{G}$. The manipulation on partition functions is based on the Fourier transformations between the space of square-integrable functions $L^2(G)$ and $L^2(\widehat{G})$ 
\begin{align}
\mathcal{F}: L^2(G)  &\rightarrow L^2(\widehat{G}) \\
f &\mapsto \mathcal{F}[f]: \chi \mapsto \int_G  f(g) \chi(g) dg \,,
\end{align}
with the inverse map $\mathcal{F}^{-1}$ given by
\begin{align}
\mathcal{F}^{-1}: L^2(\widehat{G})  &\rightarrow L^2(G) \\
\phi &\mapsto \mathcal{F}^{-1}[\phi]: g \mapsto \int_{\widehat{G}} \phi(\chi) \chi(g) d\chi \,,
\end{align}
where $dg$ denotes the Haar measure on $G$ and $d\chi$ is the correctly normalized Haar measure on $\widehat{G}$. In the case of discrete domains, the above integral descends to summation. 
\paragraph{Tannaka duality for compact nonabelian groups.} For a non-abelian compact group $G$, the set of all its finite dimensional complex representations $\text{Rep}(G)$ furnishes a monoidal category rather than a group. There is a forgetful functor (a.k.a fiber functor)
\begin{align*}
\mathfrak{F}: \text{Rep}(G)  &\rightarrow \text{Vect}_{\mathbb{C}} \\
(\pi_V, V) &\mapsto V \,,
\end{align*}
which maps the representation $(\pi_V, V)$ to its underlying vector space and homomorphism $\varphi$ between two representations $(\pi_V, V)$ and $(\pi_W, W)$ is mapped to the homomorphism between vector spaces $\mathfrak{F}(\varphi): V \longrightarrow W$. 
The set of all natural transformations from $\mathfrak{F}$ to $\mathfrak{F}$ is an algebra denoted as $\text{End}(\mathfrak{F})$. Tannaka's theorem \cite{Tannaka} states that the set of all monoidal self-conjugate natural transformations $\mathfrak{T}(G)$ is a closed subgroup of $\text{End}(\mathfrak{F})$ and $G$ is isomorphic to $\mathfrak{T}(G)$ as topological groups by the canonical map 
\begin{align}
  \pi: G  &\rightarrow \mathfrak{T}(G) \\
g &\mapsto \pi(g) \,,  \nonumber
\end{align}
note that $\pi(g)$ can be thought of as a natural transformation whose components are just the linear transformations $\pi_V(g)$ on each $V$.\\
\paragraph{Fourier transformations.} Due to the compactness of $G$, the algebra $\text{End}(\mathfrak{F})$ is simply isomorphic to $\prod_{\lambda \in G^\vee} \text{End}(V_\lambda)$ \cite{IntroTannaka}, where $G^\vee$ denotes isomorphism classes of complex irreducible representations of $G$. For $\lambda \in G^\vee$, we use $(\pi_\lambda, V_\lambda)$ to denote its representation map and the representation vector space. The dimension of $V_\lambda$ is $d_\lambda < \infty$, necessarily. 
Each $\text{End}(V_\lambda)$ is a Hilbert space endowed with a norm $\| A \|^2 := d_\lambda \text{tr}(A^* A)$, and $\prod_{\lambda \in G^\vee} \text{End}(V_\lambda)$ is made into a Hilbert space by the Hilbert sum $\sum^{\text{Hilbert}}_{\lambda \in G^\vee} \text{End}(V_\lambda)$. The Fourier transformation $\mathcal{F}$ is defined as 
\begin{align}
\mathcal{F}: L^2(G)  &\rightarrow \sum^{\text{Hilbert}}_{\lambda \in G^\vee} \text{End}(V_\lambda)\\
f &\mapsto \mathcal{F}[f]_\lambda = \int_G  f(g) \pi_\lambda(g) dg \,
\end{align} and the inverse transformation is given as
\begin{align}
\mathcal{F}^{-1}: \sum^{\text{Hilbert}}_{\lambda \in G^\vee} \text{End}(V_\lambda)  &\rightarrow L^2(G) \\
u &\mapsto \mathcal{F}^{-1}[u] : g \mapsto \sum_{\lambda \in G^\vee} \text{tr}\left( u_\lambda \pi_\lambda(g)^{-1} \right) d_\lambda \,.
\end{align}
Furthermore, if we want to restrict the transformation $\mathcal{F}$ on $L^2(Cl(G))$ (this is isomorphic to the subspace of square integrable class functions), then 
\begin{equation}
   \pi_\lambda(g) \mathcal{F}[f]_\lambda = \mathcal{F}[f]_\lambda \pi_\lambda(g) \,,
\end{equation}
for all $g \in G$ and by Schur's lemma the component map $\mathcal{F}[f]_\lambda$ must be scalar multiplication on $V_\lambda$. We denote the scalar by $\mathsf{T}[f](\lambda)$, then \begin{equation}
    \mathsf{T}[f](\lambda) = \frac{1}{d_\lambda} \int_G f(g) \chi_\lambda(g) dg  \,,
\end{equation}
where $\chi_\lambda(g) = \text{tr}\left( \pi_\lambda(g) \right)$ is the character of $(\pi_\lambda, V_\lambda)$. 
$\mathsf{T}$ defines the Fourier transformation on class functions
\begin{equation}
    \mathsf{T} :   L^2(Cl(G)) \longrightarrow L^2(G^\vee)\,,
\end{equation}
with the inverse transformation $\mathsf{T}^{-1}$ given by 
\begin{align*}
\mathsf{T}^{-1}: L^2(G^\vee)  &\rightarrow  L^2(Cl(G)) \\
\phi &\mapsto \mathsf{T}^{-1}[\phi] : g \mapsto \sum_{\lambda \in G^\vee} \phi(\lambda) \overline{\chi_\lambda(g)}  d_\lambda \,.
\end{align*}
Here $\mathsf{T}$ is the Fourier transform in~(\ref{eq:Fourier_trans}) and $\mathsf{T}^{-1}$ is its inverse.

\section{SM2: Extended Operators in Non-abelian $BF$-theory and the Induced Flavor Symmetry Action}

In this appendix we construct operators $U_\alpha(\Sigma)$ and $W_{\mathbf{R}}(\ell)$ in $BF$-theory that project to symmetry and charged operators on $\mathcal{B}_{\text{phy}}$ respectively and prove~(\ref{eq:UW_nontopo}), that is $U_\alpha(\Sigma)W_{\mathbf{R}}(\ell)$ is non-topological in the bulk $M_{d+1}$. We further prove the invariance of $\langle U_\alpha(\Sigma)W_{\mathbf{R}}(\ell) \rangle$ under the gauge and the shift transformations and comment on possible ways to obtain genuine topological actions in $BF$-theory.

\bigskip

\paragraph{The non-topological and topological operators}

We claim the following operator to be the one that leads to non-abelian $G$-flavor symmetry action after interval compactification:
\begin{equation}\label{eq:Uop}	U_\alpha(\Sigma) = \exp\left( i \int_\Sigma (\alpha, B) \right)\,.
\end{equation}
Here $B\in \Omega^{d-1}(M_{d+1},\text{ad}(P))$ where $\text{ad}(P)$ is the adjoint bundle over $M_{d+1}$ associated to the principal $G$-bundle and $\Sigma$ is a $(d-1)$-dimensional submanifold of $M_{d+1}$. The bracket $(\cdot,\cdot)$ is an $\text{ad}$-invariant inner product on $\mathfrak{g}$ such that $(a,b) = (\text{ad}_g a,\text{ad}_g b) \in \Omega^{d-1}(M_{d+1},\mathbb{R})$. We require $\alpha$ be a covariantly constant section of $\text{ad}(P)$ with respect to some flat connection $A$, i.e. there exists $A$ with $F = D_AA = 0$ such that $D_A\alpha = 0$. The non-topological nature of $U_\alpha(\Sigma)$ will be made clear in a moment.

$U_\alpha(\Sigma)$ is nevertheless gauge invariant under $A\rightarrow A' = gAg^{-1} + igdg^{-1}$, $B\rightarrow B'=g B g^{-1} \equiv \text{ad}_g B$. Under such gauge transformation we have $\alpha\rightarrow \text{ad}_g\alpha$. Therefore we have:
\begin{equation}
	U_\alpha(\Sigma) \rightarrow \exp\left( i\int_\Sigma (\text{ad}_g\alpha, \text{ad}_gB) \right) = \exp\left( i\int_\Sigma (\alpha, B) \right)
\end{equation}
due to ad-invariance of $(\cdot,\cdot)$.

The other type of operators we consider is the open Wilson line in representation $\mathbf{R}$ of $G$ stretching between a point on $\mathcal{B}_{\text{top}}$ and a local operator $\psi(x)$ on $\mathcal{B}_{\text{phy}}$ charged under $G$:
\begin{equation}
    W_\mathbf{R}(\ell) = \mathcal{P} \exp\left( i\int_\ell A_\mathbf{R} \right)\,.
\end{equation}
We note that this is an open Wilson line which is not traced.

\bigskip

\paragraph{The action of $U_\alpha(\Sigma)$ on $W_\mathbf{R}(\ell)$}

Now we are ready to prove~(\ref{eq:UW_nontopo}) in the main text. The action of $U_\alpha(\Sigma)$ on $W_\mathbf{R}(\ell)$ when $\Sigma$ links $\ell$ in $M_{d+1}$ is:
\begin{equation}\label{eq:UWshift}
	\begin{split}
		\langle U_\alpha(\Sigma)\ W_\mathbf{R}(\ell) \rangle &= \int \mathcal{D}B\ \mathcal{D}A\ \exp\left( i\int_{M_{d+1}} \text{Tr}(B\wedge F) \right)\ \exp\left( i\int_\Sigma (\alpha,B) \right)\ \mathcal{P}\exp\left( i\int_\ell A_\mathbf{R} \right) \\
		&= \int \mathcal{D}B\ \mathcal{D}A\ \exp\left( i\int_{M_{d+1}} \text{Tr}(B\wedge (F+\alpha\delta_\Sigma)) \right)\ \mathcal{P}\exp\left( i\int_\ell A_\mathbf{R} \right)
	\end{split}
\end{equation}
where in the second line $U_\alpha(\Sigma)$ is absorbed into the original $BF$-action using a $\delta$-function support $\delta_\Sigma$ over $\Sigma$. We look for a shift $A \rightarrow A-\kappa$ for which:
\begin{equation}
	F \rightarrow d(A-\kappa) - i(A-\kappa) \wedge (A-\kappa) = F-(D_A\kappa +i \kappa\wedge \kappa)\,.
\end{equation}
and require $D_A\kappa +i \kappa \wedge \kappa = \alpha \delta_\Sigma$ in order to cancel the extra $\alpha\delta_\Sigma$ in the second line of~(\ref{eq:UWshift}). Note that we have assumed that $\mathcal{D}A$ is invariant under the shift of $A$.

\begin{figure}[h]
    \centering
    \includegraphics[width=0.3\linewidth]{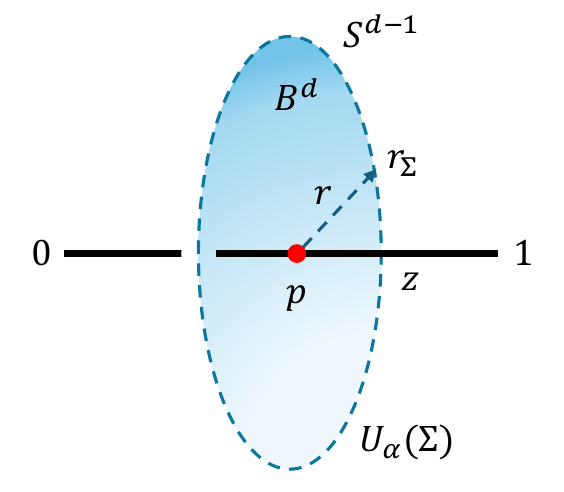}
    \caption{A local patch of $U_\alpha(\Sigma)$ linking $W_\mathbf{R}(\ell)$ near $p \in [0,1]$.}
    \label{fig:UW}
\end{figure}

To simplify~(\ref{eq:UWshift}) we let $\Sigma$ be the boundary $(d-1)$-sphere $S^{d-1}$ of a $d$-ball $B^d$ with infinitesimal radius $r_\Sigma$ that intersects $\ell$ transversally at $p$, see Figure~\ref{fig:UW}.
We place $p$ at the center of $B^d$ and define the following differential form with $\delta$-function support:
\begin{equation}
	\mathscr{H}(r,\Sigma) = H(r-r_\Sigma) \delta(z-p) dz,\ H(r-r_\Sigma) = \begin{cases}
		0,\ r > r_\Sigma \\
		1,\ r \leq r_\Sigma
	\end{cases}
\end{equation}
where $z$ is local coordinate describing the direction in $M_{d+1}$ normal to $B^d$ and $r$ is radial direction of $B^d$. Clearly we have $d\mathscr{H}(r,\Sigma) = \delta(r-r_\Sigma)\delta(z-p)drdz \equiv \delta_\Sigma$. Let $\kappa = \alpha\mathscr{H}(r,\Sigma)$, we have:
\begin{equation}
	D_A\kappa = (D_A\alpha) \mathscr{H}(r,\Sigma) + \alpha \delta_\Sigma\ \text{and}\ \kappa\wedge \kappa \propto dz\wedge dz = 0\,.
\end{equation}
Thus under $A\rightarrow A - \kappa$ we have:
\begin{equation}\label{eq:shrink_rSigma}
	\begin{split}
		\int_{M_{d+1}} \text{Tr}(B\wedge (F + \alpha\delta_\Sigma)) \rightarrow \int_{M_{d+1}} \text{Tr}(B\wedge (F - (D_A\alpha) \mathscr{H}(r,\Sigma))) &= S_{BF} - \int_{M_{d+1}} \text{Tr}(B\wedge D_A\alpha) \mathscr{H}(r,\Sigma) \\
		&= S_{BF} - \int_{M_d} \text{Tr}(B\wedge D_A\alpha) H(r-r_\Sigma)
	\end{split} 
\end{equation}
where $M_d = M_{d+1}|_{z = 0}$. Since $\int_{M_d} \text{Tr}(B\wedge D_A\alpha) H(r-r_\Sigma) \propto r_\Sigma^{d}$ (for well-behaved $B$ and $\alpha$),~(\ref{eq:UWshift}) becomes:
\begin{equation}\label{eq:UW_VEV}
    \langle U_\alpha(\Sigma)\ W_\mathbf{R}(\ell) \rangle = \int \mathcal{D}B\ \mathcal{D}A\ \exp\left( i\int_{M_{d+1}} \text{Tr}(B\wedge F) \right)\ \mathcal{P}\exp\left( i\int_\ell (A_\mathbf{R} - \kappa_\mathbf{R}) \right) = \langle \mathcal{P} \exp\left( i\int_\ell (A_\mathbf{R} - \kappa_\mathbf{R}) \right)\rangle
\end{equation}
in the $r_\Sigma\rightarrow 0$ limit. Since~(\ref{eq:UW_VEV}) holds also as an operator equation, we will drop the outmost angle bracket from now on.

The path-ordered integral on the RHS of~(\ref{eq:UW_VEV}) can be evaluated using geometric integration to be~\cite{grossman1972non}:
\begin{equation}\label{eq:UW_geo_int}
    \begin{split}
        U_\alpha(\Sigma) W_\mathbf{R}(\ell) &= \lim_{\Delta z \rightarrow 0}\prod_{z_i} \exp\left( i (A_\mathbf{R}(z_i) - \kappa_\mathbf{R}(z_i)) \Delta z \right) \\
        &= \lim_{\Delta z \rightarrow 0} \prod_{z_i > p} \exp\left( i A_\mathbf{R}(z_i) \Delta z \right) \times \exp\left( -i\alpha_\mathbf{R}(z)\delta(z-p) \Delta z \right) \times \prod_{z_i < p} \exp\left( i A_\mathbf{R}(z_i) \Delta z \right) \\
        &= \mathcal{P} \left[ \exp\left( i \int_p^1 A_\mathbf{R} \right) \exp\left(-i\alpha_\mathbf{R}(p)\right) \exp\left( i \int_0^p A_\mathbf{R} \right) \right]
    \end{split}
\end{equation}
for a partition $\{z_i\}$ of $\ell\cong[0, 1]$ and $\Delta z = \max_i |z_{i+1} - z_i|$. Therefore~(\ref{eq:UW_nontopo}) is proved. It is immediately seen from~(\ref{eq:UW_geo_int}) that $U_\alpha(\Sigma) W_\mathbf{R}(\ell)$ cannot be topological since the RHS depends explicitly on $p$. This implies that $U_\alpha(\Sigma)$ itself cannot be topological since $W_\mathbf{R}(\ell)$ is and it is impossible that the linking of two topological operators leads to a non-topological outcome.

Fixing $z_p = 0$,~(\ref{eq:UW_geo_int}) becomes:
\begin{equation}\label{eq:UW_p=0}
    U_\alpha(\Sigma) W_\mathbf{R}(\ell) = \mathcal{P} \left[ \exp\left( i \int_0^1 A_\mathbf{R} \right) \exp\left(-i\alpha_\mathbf{R}(0)\right) \right] = W_\mathbf{R}(\ell) e^{-i\alpha_\mathbf{R}(0)}\,.
\end{equation}
By the construction illustrated in Figure~\ref{fig:U_on_boundary} of the main text, after an interval compactification on $[0,1]$,~(\ref{eq:UW_p=0}) leads to the sought-after non-abelian $G$-flavor symmetry action on a local operator $\psi(x)$ at $z = 1$.

\bigskip

\paragraph{Gauge invariance of $\langle U_\alpha(\Sigma)\ W_\mathbf{R}(\ell) \rangle$.}

Though the invariance of $\langle U_\alpha(\Sigma)\ W_\mathbf{R}(\ell) \rangle$ under $A\rightarrow A' = gAg^{-1} + i gdg^{-1}$, $B\rightarrow B' =g B g^{-1} \equiv \text{ad}_g B$ is immediate, it is well-known that $BF$-theory also enjoys a shift gauge symmetry $B\rightarrow B' = B + D_AK$ for $K\in\Omega^{n-2}(M_{d+1},\mathfrak{g})$ hence we would like to check if $\langle U_\alpha(\Sigma)\ W_\mathbf{R}(\ell) \rangle$ is invariant under such shift (with suitable $K$). To this end, we will check if the following equation holds:
\begin{equation}
	\begin{split}
		\langle U_\alpha(\Sigma)\ W_\mathbf{R}(\ell) \rangle &=\int \mathcal{D}B\ \mathcal{D}A\ \exp\left( i\int_{M_{d+1}} \text{Tr}(B\wedge F) \right)\ \mathcal{P}\exp\left( i\int_\ell (A_\mathbf{R} - \kappa_\mathbf{R}) \right) \\
        &= \int \mathcal{D}B\ \mathcal{D}A\ \exp\left( i\int_{M_{d+1}} \text{Tr}((B+D_AK)\wedge F) \right)\ \mathcal{P}\exp\left( i\int_\ell (A_\mathbf{R} - \kappa_\mathbf{R}) \right)\,.
	\end{split}
\end{equation}
A necessary condition for the above equation to hold is to require:
\begin{equation}
	\int_{M_{d+1}} \text{Tr}(D_AK\wedge F) = 0\,.
\end{equation}
Due to Bianchi identity $D_AF = 0$, we have:
\begin{equation}
	\int_{M_{d+1}} \text{Tr}(D_AK\wedge F) = \int_{M_{d+1}} d \text{Tr}(K\wedge F) = \int_{\partial M_{d+1}} \text{Tr}(K\wedge F)\,.
\end{equation}
Requiring $K|_{\partial M_{d+1}} = 0$ implies that $\int_{M_{d+1}} \text{Tr}(D_AK\wedge F) = 0$ hence $\langle U_\alpha(\Sigma)\ W_\mathbf{R}(\ell) \rangle$ is invariant under $B\rightarrow B' = B + D_AK$.

\bigskip

\paragraph{Topological action in $BF$-theory.}

It is clear that $U_\alpha(\Sigma) W_{\mathbf{R}}(\ell)$ fails to be topological due to its explicit dependence on $p$. Therefore, in order obtain a topological action one may try to remove the $p$-dependence from~(\ref{eq:UW_geo_int}).

The most economic way to do so is to let $e^{-i\alpha_\mathbf{R}(p)}$ commutes with all elements of $G$ so that $e^{-i\alpha_\mathbf{R}(p)}$ in the middle of the path-ordered exponential can be factored out. This restricts $e^{-i\alpha_\mathbf{R}(p)}$ to be a constant in the center $Z(G)$ of $G$. In this case $U_\alpha(\Sigma)$ generates nothing but the 1-form $Z(G)$-symmetry of $BF$-theory.

Another way to remove $p$-dependence is to define an operator $\widetilde{U}_{\widetilde{\alpha}}(\Sigma)$ for a fixed flat connection $A$:
\begin{equation}\label{eq:UopTilde}
    \widetilde{U}_{\widetilde{\alpha}}(\Sigma) = \int dg \exp\left( i\int_\Sigma(\widetilde{\alpha}, B) \right)
\end{equation}
where $dg$ is the Haar measure of $G$ and $\widetilde{\alpha}$ is the parallel transport of $g\alpha(p)g^{-1}$ for a covariantly constant $\alpha$ at $p$ by the holonomy of $A$. The above definition is independent of the choice of $p$. Unlike $\alpha$, we allow $\widetilde{\alpha}$ to be defined up to conjugation. In other words, $e^{-i\widetilde{\alpha}}\in Cl(G)$. The action of $\widetilde{U}_{\widetilde{\alpha}}(\Sigma)$ on $W_\mathbf{R}(\ell)$ is:
\begin{equation}\label{eq:to_prove}
    \widetilde{U}_{\widetilde{\alpha}}(\Sigma) W_\mathbf{R}(\ell) = \mathcal{P}\left( e^{i\int_p^1 A_\mathbf{R}}\ \left(\int dg\ g e^{-i\alpha_\mathbf{R}(p)}g^{-1} \right)\ e^{i\int_0^p A_\mathbf{R}} \right)
\end{equation}
where $dg$ is the Haar measure of $G$. For irreducible $\mathbf{R}$, we have by Schur's lemma~\cite{Cordova:2022rer}:
\begin{equation}
    \int d\mu\ g e^{-i\alpha_\mathbf{R}(p)} g^{-1} = \frac{\chi_\mathbf{R}(e^{-i\alpha(p)})}{\dim \mathbf{R}} \times \mathbb{I}_{\dim \mathbf{R} \times \dim \mathbf{R}}\, 
\end{equation}
where the group volume is normalized as $\int d\mu=1$. Therefore, after taking VEV~(\ref{eq:to_prove}) becomes:
\begin{equation}
    \langle \widetilde{U}_{\widetilde{\alpha}}(\Sigma) W_\mathbf{R}(\ell) \rangle = \frac{\chi_\mathbf{R}(e^{-i\alpha(p)})}{\dim \mathbf{R}} \langle W_\mathbf{R}(\ell) \rangle
\end{equation}
the RHS of which is manifestly independent of $p$ since $e^{-i\alpha(p)}$ at different $p$ all live in the same conjugacy class of $G$.

The operator $\widetilde{U}_{\widetilde{\alpha}}(\Sigma)$ can be expressed alternatively as
    \begin{equation} 
        \widetilde{U}_{\widetilde{\alpha}}(\Sigma) = \int \mathcal{D}g(\Sigma) \mathcal{D} \varphi(\Sigma) \exp \left(i \int_{\Sigma} (\textrm{ad}_{g} \alpha , B + D_A \varphi) \right)
    \end{equation}
where $\alpha$ is a $\mathfrak{g}$-valued constant and we introduce two auxiliary fields, $G$-valued function $g(\Sigma)$ and ad($P$)-valued $(d-2)$-form $\varphi(\Sigma)$, both supported on $\Sigma$.  This formulation is manifestly invariant under the gauge transformations $A\rightarrow gAg^{-1} + i gdg^{-1}$ and $B\rightarrow g B g^{-1}$ if we assume the measure $\mathcal{D}g$ and $\mathcal{D\varphi}$ are invariant. It is also invariant under the gauge transformation $B\rightarrow B+D_AK$ since one can absorb $K$ into the auxiliary field $\varphi$. The equivalence between the two formulations can be shown as follows. Integrating out $\varphi$ yields a delta function $\delta(D_A(\textrm{ad}_g \alpha))$ on $\Sigma$. It restricts $g(\Sigma)$ to configurations where $\textrm{ad}_g \alpha$ is a covariantly constant section with respect to a fixed $A$. Further integrating over $g(\Sigma)$ sums over all such covariantly constant sections and one recovers \eqref{eq:UopTilde} up to a normalization constant.

\section{SM3: $BF$-Theory as SymTFT from Holography}

It is interesting to look for other approaches leading to $BF$-action as the bulk SymTFT action that governs the 0-form global $G$-symmetry of the physical theory $\mathcal{T}_G$ on the boundary. It was pointed out in~\cite{Bonetti:2024cjk} that a $BF$-action can be obtained on the boundary of AdS$_{d+1}$ with Yang-Mills theory in the bulk. The derivation requires decoupling of a kinetic term $B\wedge \star B$ for an auxiliary field $B$ from the Yang-Mills action on AdS$_{d+1}$ written in the first-order $BFYM$ form~\cite{Cattaneo:1995xa, Martellini:1996ek, Cattaneo:1997eh}. However, naively taking the $g\rightarrow 0$ limit of the $BFYM$ action results in the following action~\cite{Cattaneo:1997eh} :
\begin{equation}
	S = \int \Tr B\wedge F + \frac{1}{2} D\eta \wedge \star D\eta 
\end{equation}
with St\"uckelberg field $\eta$ to account for the correct number of degrees of freedom rather than the $BF$-action. Therefore one has to be careful when deriving the bulk $BF$-action via holography.  In this section, we will derive that the SymTFT of $\mathcal{T}_G$ is a $BF$-theory using AdS/CFT more rigorously.

Our derivation relies on one of the earliest observations in AdS/CFT that when AdS$_{d+1}$ theory has a gauge group $G$, the group $G$ is a global symmetry group of the theory on the boundary and the currents of the boundary theory couple to the bulk gauge fields as $A_a J^a$, $a = 1,\cdots, \dim(G)$~\cite{Maldacena:1997re, Witten:1998qj}, in other words there is the following term in the generating function:
\begin{equation}\label{eq:AJ_coupling}
	\int d^d x\ A_\mu J^\mu\,.
\end{equation}
In the absence of anomalies, the bulk theory must be invariant under gauge transformation $\delta A = D\lambda$, thus the above boundary coupling must be invariant as well, i.e. we have:
\begin{equation}\label{eq:general_bulk_boundary_current}
	0 = \delta \int d^d x A_\mu J^\mu = - \int d^d x\  \lambda D_\mu J^\mu
\end{equation}
for arbitrary $\lambda$. Therefore the current $J^\mu$ on the boundary is covariantly conserved in the presence of the gauge field $A$, which becomes a conserved current when $A$ is totally decoupled.

To obtain~(\ref{eq:AJ_coupling}), we start with the Yang-Mills action in AdS$_{d+1}$ and focus on $d = 4$ for concreteness:
\begin{equation}
	S_{YM} = \int d^5x \ \sqrt{-g} \left( - \frac{1}{4} F_{\mu\nu}F^{\mu\nu} \right)\,.
\end{equation}
The equations of motion for $S_{YM}$ are:
\begin{equation}\label{eq:EOM}
	\begin{split}
		\eta^{ij} \left( \partial_i \partial_r A_l -i \left[A_i,\partial_r A_l \right] \right) &= 0\, , \\
            \left(\frac{3}{r} + \partial_r \right) \partial_r A_l + \frac{L^4}{r^4}\eta^{ik} \left(\partial_i F_{kl} -i \left[A_i,F_{kl} \right]\right) &= 0\, ,
	\end{split}
\end{equation}
where the AdS$_5$ boundary is at $r\rightarrow \infty$ and $L$ is the AdS length scale. In the gauge $A_r=0$, the solution to the EOM can be expanded as:
\begin{equation}
    A_i(\vec{x},t,r) = \alpha_i(\vec{x},t) L + \gamma_i (\vec{x},t)\frac{L^5 \log r}{r^2} + \beta_i(\vec{x},t)\frac{L^5}{r^2} + \cdots
\end{equation}
similar to what has been done in~\cite{Klebanov:1999tb} for scalar field and in~\cite{Marolf:2006nd, DeWolfe:2020uzb} for abelian gauge field. Here $(\vec{x},t)$ is the coordinate of the AdS boundary and the fields $\alpha_i,\beta_i,\gamma_i$ are all Lie-algebra valued and $\alpha_i$ and $\beta_i$ are independent modes. The first equation of~(\ref{eq:EOM}) implies that $\beta_i$ should satisfy
\begin{equation}
    \eta^{ij} \left( \partial_i \beta_j -i L\left[{\alpha_i},\beta_j\right]\right) = 0
\end{equation}
to the leading order in $r$. We define the 4D connection and the field strength as follows
\begin{equation}\label{eq:define_fieldstrength}
    a_i(\vec{x},t) = L \alpha_i(\vec{x},t),\quad f_{ij}(\vec{x},t) = \partial_i a_j - \partial_j a_i -i \left[a_i , a_j \right]
\end{equation}
and the covariant derivative $\mathcal{D}_i = \partial_i - i\left[a_i,\cdot \right]$. One can show that $\beta_i$ must satisfy the following equation:
\begin{equation}\label{eq:beta_EOM}
        \eta^{ij} \mathcal{D}_i \beta_j = \mathcal{D}^i \beta_i = 0.
\end{equation}
In other words, we have $\mathcal{D}\star\beta = 0$ where $\beta := \beta_i dx^i$ is a 1-form on the AdS$_5$ boundary. From the first equation of~(\ref{eq:EOM}), we can further solve $\gamma_i$ as:
\begin{equation}
	\gamma_i (\vec{x},t) =  \frac{1}{2L} \eta^{jk} \mathcal{D}_j f_{ki}
\end{equation}
which means that $\gamma_i$ is not an independent mode.

Now we are ready to study the variation of $S_{YM}$ with respect to $A$, i.e. $\delta S_{YM}$. For this variation, we have:
\begin{equation}
    \begin{split}
        \delta S_{YM} &= \delta_A \left(\int d^5 x\ \sqrt{-g} \left[-\frac{1}{4} \Tr F_{\mu \nu} F^{\mu \nu} \right]\right) \\
        &= \int d^5 x\ \sqrt{-g} \left( -\frac{1}{2} \Tr \left(\partial_{\mu} \delta A_{\nu} - \partial_{\nu} \delta A_{\mu} -i \delta A_{\mu} A_{\nu} -i A_{\mu} \delta A_{\nu} +i \delta A_{\nu} A_{\mu} +i A_{\nu} \delta A_{\mu}\right)F^{\mu \nu} \right) \\
        &= \int d^5 x\  \partial_{\mu} \left(\sqrt{-g} \left[-\textrm{Tr} \delta A_{\nu} F^{\mu \nu} \right]\right) + \int d^5 x \textrm{Tr} \delta A_{\nu} D_{\mu} F^{\mu \nu} \\
        &= -\frac{R^3}{L^3} \int d^4x \Tr \delta A_i F_{rj} g^{rr} g^{ij} 
    \end{split}
\end{equation}
where in the last step we use equations of motion in the bulk and Stokes' theorem to turn the bulk action into an action on the boundary of AdS$_5$ at large cut-off $R$ of $r$-direction. Here the $r$-index is not summed over. Evaluating $\delta S_{YM}$ at the solution to~(\ref{eq:EOM}), we have:
\begin{equation}\label{eq:variation_S}
	\begin{split}
		\delta S_{YM} &= 2L^3 \int d^4x \Tr \delta \alpha_i \beta^i - L^3(1-2 \log R) \int d^4x \Tr \delta \alpha_i \gamma^i \\
		&= 2L^3 \int d^4x \Tr \delta \alpha_i \beta^i -\frac{L^2}{2}(1-2 \log R) \int d^4x \Tr \delta \alpha^i \mathcal{D}^j f_{ji} \\
		&= 2L^3 \int d^4x \Tr \delta \alpha_i \beta^i +\frac{L^2}{2}(1-2 \log R) \int d^4x \Tr  \left( \mathcal{D}^j \delta \alpha^i \right) f_{ji} \\
		&= 2L^3 \int d^4x \Tr \delta \alpha_i \beta^i +\frac{L}{2}(1-2 \log R) \int d^4x \Tr  f_{ij} \delta f^{ij}
	\end{split}
\end{equation}
where in the last step we use the fact that $\delta f = \mathcal{D}\delta a$ and $a = L\alpha$ as already stated in~(\ref{eq:define_fieldstrength}). The second term of~(\ref{eq:variation_S}) which diverges logarithmically as $R\rightarrow \infty$, is to be cancelled by the variation of the following counter term on the AdS boundary:
\begin{equation}
	S_{\text{ct}} = L \int d^4x  \left( -\frac{1}{g^2(R)}\Tr f_{ij}f^{ij} \right)
\end{equation}
for the choice $\frac{1}{g^2(R)} = \frac{1}{2} - \log R$ (see~\cite{DeWolfe:2020uzb} for treatment of abelian field). Such a choice of $S_{\text{ct}}$ ensures that both $S_{YM}+S_{\text{ct}}$ and $\delta(S_{YM}+S_{\text{ct}})$ are finite. In particular, we have:
\begin{equation}\label{eq:variation_Stot}
	\delta S_{\text{tot}} = \delta(S_{YM}+S_{\text{ct}}) = 2L^3 \int d^4x \Tr \delta\alpha_i\beta^i\,.
\end{equation}
Therefore, the response of the source $\alpha_i$ on the boundary is:
\begin{equation}\label{eq:variation_current}
	\langle J^i(\vec{x},t) \rangle = \frac{\delta S_{\text{tot}}}{\delta \alpha_i} = 2L^3 \beta^i\,.
\end{equation} 
By~(\ref{eq:beta_EOM}) it is immediate to see that $J^i(\vec{x},t)$ is a covariantly conserved current on the AdS boundary when coupled to background $G$-gauge field $\alpha_i$. By the argument around~(\ref{eq:general_bulk_boundary_current}), we conclude that the boundary theory must have global $G$-symmetry.

We note that the above derivation leading to~(\ref{eq:variation_Stot}) is totally in parallel with the derivation in~\cite{DeWolfe:2020uzb} for $U(1)$ gauge theory in AdS$_5$ and the consequential $U(1)$ current on the AdS boundary. The main difference is that in the abelian case we have $DJ = dJ = 0$ hence the current is not only covariantly conserved, but is indeed conserved.

Looking back at the derivation leading to~(\ref{eq:variation_current}), it is clear in retrospect that what really matters is the existence of a covariantly conserved current as the response of varying a source in certain boundary action~(\ref{eq:variation_Stot}). Therefore one may ask if there exists a different bulk theory the variation of whose action is identical to $\delta S_{\text{tot}}$. Such a bulk theory would lead to the same covariantly conserved current on the boundary and hence is indistinguishable from Yang-Mills on AdS if we care about only the global symmetry of the boundary theory. Moreover, such a theory does not have to be defined on AdS as long as the boundary of the space on which it is defined is isomorphic to the AdS boundary.

Let us stick with the above AdS$_5$ example and consider a non-abelian $BF$-theory defined on $M_4\times(0,1]$ homeomorphic to the neighborhood of AdS$_5$ boundary which is isomorphic to $M_4\times \{1\}$. For this $BF$-theory, we have:
\begin{equation}
	\begin{split}
		\delta S = \delta\left( \int \Tr B\wedge F\right) = \int d^5x \Tr \left(\delta \widetilde{B}^{\mu\nu}F_{\mu\nu} + \widetilde{B}^{\mu\nu} D_\mu\delta A_\nu\right) = \int d^5x \Tr \left(\delta \widetilde{B}^{\mu\nu} F_{\mu\nu} - D_\mu \widetilde{B}^{\mu\nu} \delta A_\nu + \partial_\mu (\widetilde{B}^{\mu\nu}\delta A_\nu)\right)
	\end{split}
\end{equation}
where $\widetilde{B} = * B$ is the Hodge dual of $B$. Applying the EOM of $A$ and $B$ (or $\widetilde{B}$) similar to what we did in the derivation of~(\ref{eq:variation_current}), we have:
\begin{equation}\label{eq:variation_BF}
	\delta S = \int d^5x\ \partial_\mu \Tr (\widetilde{B}^{\mu\nu} \delta A_\nu) = \int d^4x \Tr \widetilde{B}^{ri} \delta A_i \,.
\end{equation}
Therefore, the variation of $S$ with respect to $A_i$ results in a current on $M_4$:
\begin{equation}
	\langle J^i(\vec{x},t) \rangle = \frac{\delta S}{\delta A_i} = \widetilde{B}^{ri}
\end{equation}
and the EOM $D_i \widetilde{B}^{ri} = 0$ ensures that $J$ is covariantly conserved.

Comparing~(\ref{eq:variation_BF}) with~(\ref{eq:variation_Stot}), it is immediate to see that the variation of $BF$-action and that of Yang-Mills action near $M_4$ both lead to identical physical consequences, i.e. the existence of a covariantly conserved current on $M_4$ as the response of variation. Rather than taking the weak coupling limit of Yang-Mills or $BFYM$ on AdS, we obtain $BF$-action without auxiliary $\eta$ field as the theory whose variation is identical to~(\ref{eq:variation_Stot}) hence leading to the same current on the boundary.  We claim that it is in this sense that $BF$-theory arises as the SymTFT in the bulk that governs the global $G$-symmetry of the physical theory on the boundary. This concludes our holographic derivation of $BF$-theory being the SymTFT of global $G$-symmetry.

\end{document}